\documentclass[epj]{svjour}
\usepackage{amssymb}
\usepackage{amsmath}
\usepackage[english]{babel}  
\usepackage{graphicx}
\usepackage{wasysym}   
\usepackage{subcaption}    

\renewcommand{\Re}{\operatorname{Re}}
\renewcommand{\Im}{\operatorname{Im}}
\begin{document}
\title{Nonlocal control of pulse propagation in excitable media}
\author{Clemens A. Bachmair \and Eckehard Sch\"{o}ll
}                     
\institute{Insitut f\"{u}r Theoretische Physik, Technische Universit\"{a}t Berlin, 10623 Berlin, Germany}
\date{Received: date / Revised version: date}

\abstract{
We study the effects of nonlocal control of pulse propagation in excitable media. As a generic example for an 
excitable medium the FitzHugh-Nagumo model with diffusion in the activator variable is considered. Nonlocal 
coupling in form of an integral term with a spatial kernel is added. We find that the nonlocal coupling modifies 
the propagating pulses of the reaction-diffusion system such that a variety of spatio-temporal patterns are generated including 
acceleration, deceleration, suppression, or generation of pulses, multiple pulses, and blinking pulse trains. It is shown that one can observe these effects for various choices of the integral kernel and the coupling 
scheme, provided that the control strength and spatial extension of the integral kernel is appropriate. In 
addition, an analytical procedure is developed to describe the stability borders of the spatially homogeneous 
steady state in control parameter space in dependence on the parameters of the nonlocal coupling. 
\PACS{05.45.-a, 05.65.+b, 82.40.Ck}
} 

\maketitle

\section{Introduction}
\label{intro}
Excitable media occur in a wide range of physical, chemical, biological, as well as socio-economic systems, and they are often modelled as nonlinear reaction-diffusion systems \cite{HAK83,KUR84,MIK94,KAP95a,KEE98}, which
for certain parameter ranges support traveling excitation pulses. 
An important application in neuroscience is the propagation of information as electrical pulses along a nerve fiber.
Excitation pulses in the form of spreading depression or depolarization are also associated with pathological states
of brain activity occurring, for instance, during migraine or stroke \cite{KAR13,DRE11,DAH09a}. 
If, in addition to local diffusion, nonlocal spatial coupling is present, this represents an important internal control mechanism 
in neuronal wave dynamics. The resulting spatio-temporal patterns in nonlocally coupled reaction-diffusion systems can be quite complex, ranging from 
traveling waves, Turing patterns, and pulse trains to spatio-temporal chaos
\cite{BEL81,MAZ97,SHE97a,KUR98,HIL01,LI01a,NIC02,VAR05,NIC06,BOR06,GEL10,COL14,GEL14,LOE14}.
Such nonlocal couplings in the form of integrals with a spatial kernel can be derived as limiting cases of two- or three-component activator-inhibitor reaction-diffusion models with or without advection, when fast inhibitor variables are eliminated adiabatically
\cite{PIS06,PET94,NIC02,TAN03,SHI04,SIE14}. In particular, asymmetric spatial kernels \cite{SIE14} arise from differential advection of some chemical species, e.g., in heterogeneous catalysis, marine biology, or ecology \cite{MID92a,YOC10a,ROV93,KHA95b,MAL96a,HAR01a}. It has been shown that the adiabatic elimination of a fast variable in a three-variable system results in a nonlocal integral term that has the form of an exponential kernel \cite{SHI04}, see also \cite{SIE14}.
In the limit of global spatial coupling, i.e., fast diffusion of the eliminated variable, a large number of experimental and theoretical studies, e.g., in the CO oxidation on platinum surfaces \cite{BAE94,BET04a} and other catalytic processes \cite{MID92a}, as well as in electrochemistry \cite{PLE01} or in semiconductors \cite{MEI00b,SCH01} have shown that global feedback can control propagating waves and generate spatially periodic patterns such as Turing patterns or travelling waves \cite{MIK06}. 
In electrochemistry, models with explicit nonlocal coupling have been derived from more elementary models by applying a Green's function formalism, see e.g. \cite{CHR02a}.
Nonlocal coupling plays also an essential role in the formation of chimera states in systems of coupled oscillators \cite{KUR02a,ABR04,HAG12,TIN12,MAR13,OME13,ZAK14,PAN14}. 

In neuronal systems, especially in the visual cortex, several experimental studies have given evidence for nonlocal long-range connectivity of neurons \cite{KAN02,XU04b}.
In the framework of the FitzHugh-Nagumo model of excitable dynamics augmented by a spatially discrete nonlocal coupling term, it was shown that traveling pulses in one spatial dimension can be suppressed by various control schemes of spatially nonlocal or time-delayed coupling \cite{DAH08,SCH09c}. This has been explained by the effective change of the excitation threshold of the original reaction-diffusion system by nonlocal interaction at a certain spatial distance and a certain coupling strength. 
The nucleation and propagation of spatially localized
reaction-diffusion waves has also been studied in two-dimensional flat \cite{DAH12b} and curved surfaces \cite{KNE14}
under global spatial feedback in the framework of the FitzHugh-Nagumo model. In
particular, it was shown that the stability of propagating wave segments depends crucially on the curvature of the
surface \cite{KNE14}. In two- and three-dimensional excitable media more complex spatio-temporal patterns like 2D spiral waves and 3D scroll waves can arise \cite{WIN87,ALO03,DAE13}.


While much work has been done on the effect of time-delayed feedback upon spatio-temporal patterns in reaction-diffusion systems, e.g.,
\cite{AHL07,KYR09,GUR13b,STI13}, no systematic investigation of the influence of a distributed nonlocal spatial coupling upon pulse propagation in excitable media has been carried out to the best of our knowledge.

In this paper we restrict ourselves to one-dimensional excitable media modelled by the FitzHugh-Nagumo equations, and study the effects of nonlocal spatial couplings upon pulse propagation. In contrast to previous work \cite{DAH08,SCH09c}, where the nonlocal coupling was modelled by the dynamical variable taken at a certain discrete spatial distance, here the nonlocal coupling is described by an integral over space weighted with a spatial kernel, which is chosen as rectangular, exponential, or Mexican-hat like. A Mexican-hat kernel is a superposition of two Gaussians with opposite sign and different widths, and hence models a nonmonotonic spatial interaction which is attractive for small distances and repulsive for large distances; it is of particular relevance for interacting cortical neurons since they combine excitatory coupling of neighbouring cells with long-range inhibitory interactions of distant cells \cite{KAN02,HUT03,ZHA14}.
We show that the resulting spatio-temporal patterns can be acceleration, deceleration, and suppression of propagating pulses as well as generation of Turing patterns, and multiple pulses and blinking traveling waves, depending upon the type of spatial kernel, the coupling strength, the coupling range, and the coupling scheme. Specifically, we will extend our previous work \cite{SCH09c} in three directions: First of all, we consider spatially extended integral kernels, such as the exponential and the Mexican hat function. Second, we do not only investigate the suppression of propagating pulses but show that nonlocal control can also be used to accelerate or decelerate pulses as well as to generate pulses or wave trains. We will take all these effects into account and explore systematically the control parameter space showing where to expect which kind of behavior, i.e., we vary the strength and range of the coupling, and the coupling scheme for different nonlocal coupling kernels. Thus we provide an exhaustive picture how nonlocal coupling affects pulse propagation in an excitable medium.
Finally, we will gain some analytical insights by analyzing the stability of the homogeneous steady state, and determining how its stability is affected by nonlocal control. We show that the control scheme enables the generation of a plethora of different instabilities.

The organization of the paper is as follows.
In section 2 we introduce the FitzHugh-Nagumo model with nonlocal coupling. In section 3 a linear stability analysis of the homogeneous steady state is performed, which gives rise to various spatio-temporal instabilities. In section 4 we present simulations of the nonlinear reaction-diffusion equations with different nonlocal kernels, and discuss the resulting complex patterns.

\section{FitzHugh-Nagumo model with nonlocal coupling}
\begin{figure}
\resizebox{.5\textwidth}{!}{%
  \includegraphics[width=0.2\textwidth]{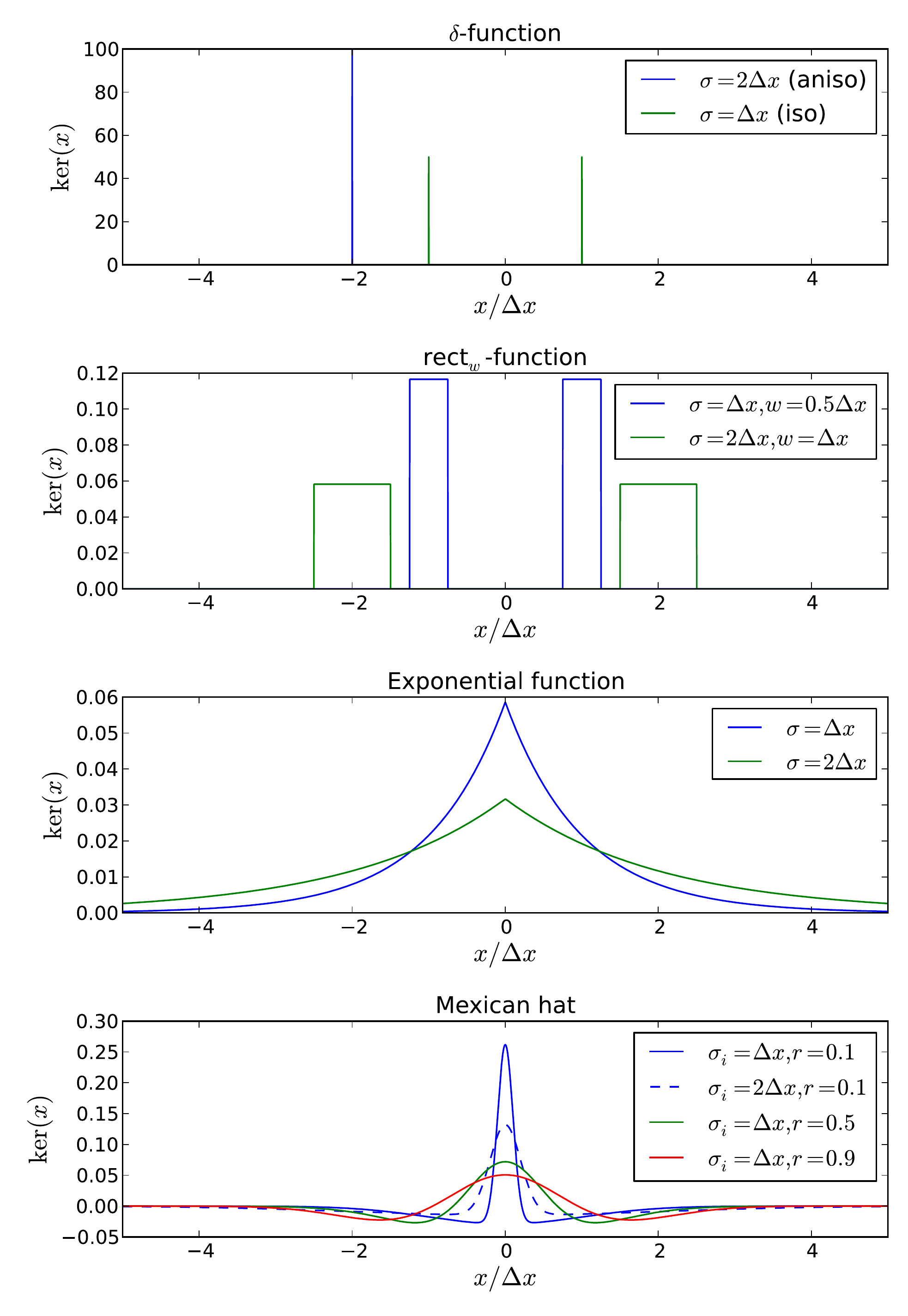}
}
\caption{Various nonlocal control kernels $\ker\left(x\right)$, see Eqs.~(\ref{eq:control_term}),(\ref{eq:kerDelta})-(\ref{eq:kerMexhat}), for different characteristic interaction lengths $\sigma$. The length is scaled in units of $\Delta x=8.59$ which is the pulse width of the uncontrolled pulse for $\left(\varepsilon,\beta\right)=\left(0.08,1.2\right)$. The anisotropic $\delta$-function (top panel, labeled {\em aniso}) corresponds to backward coupling for a pulse traveling in positive $x$-direction.}
\label{fig:kernels}
\end{figure}

In this paper an excitable medium is modeled by the generic FitzHugh-Nagumo (FHN) system \cite{FIT61,NAG62,LIN04} which is spatially extended in one dimension by diffusion only in the activator variable:
\begin{eqnarray}
\dot u =& u - \frac{u^3}{3} - v + \partial_{xx}u,\label{eq:1a}\\
\epsilon^{-1} \dot v =& u + \beta,\label{eq:1b}
\end{eqnarray}
where $u$ and $v$ denote the activator (membrane potential) and the inhibitor (recovery variable), respectively,  
$0 < \varepsilon \ll 1$ separates the time scale of the fast activator $u$ and the slow inhibitor $v$, and $\beta$ is an indicator for the excitability of the system. The diffusion constant is scaled to unity. In the excitable regime ($\beta > 1$) there exists a stable homogeneous steady state. These equations show -- for appropriate values of $\varepsilon$ and $\beta$ -- the well-known behavior of supporting traveling pulses and waves after supra-threshold excitations. Throughout this paper we use the parameters $\varepsilon=0.08$
and $\beta= 1.2$. The spatially homogeneous steady state is given by:
\begin{eqnarray}
\left(u^*, v^* \right)^T = \left(-\beta, \frac{\beta^3}{3} - \beta \right)^T.\label{eq:fixed_point}
\end{eqnarray}
Now a nonlocal control term $\hat K$ is added to the system Eqs.~\eqref{eq:1a}~and~\eqref{eq:1b}. Introducing the vector notation 
$U= \left(u, v \right)^T$ we can write the nonlocally coupled FHN system as:
\begin{eqnarray}
E \dot U = F\left(U\right) +\hat D U + \hat K U,\label{eq:main}
\end{eqnarray}
where 
\begin{equation}\label{eq:couplingMatrixE}
E = \begin{pmatrix} 1 & 0 \\ 0 & \varepsilon^{-1}  \end{pmatrix}\,,
\end{equation}
is the time scale separation matrix, 
$F\left(U\right)$ describes the FHN dynamics given by the right hand sides of Eqs.~(\ref{eq:1a}),(\ref{eq:1b}), and
\begin{equation}\label{eq:couplingMatrixI}
\hat D = \begin{pmatrix} \partial_{xx} & 0 \\ 0 & 0  \end{pmatrix}\,,
\end{equation}
is the diffusion matrix. The nonlocal control operator $\hat K$ is defined as follows:
\begin{eqnarray}
\hat K U = \kappa A \left[ \int U\left(x-x'\right) \ker\left(x'\right)dx'- \eta U\left(x\right)\right],\label{eq:control_term}
\end{eqnarray}
where $\kappa\in\mathbb R$ is the control strength, $A\in\mathbb{R}^{2\times2}$ is a $2 \times 2$ coupling matrix, which 
may be chosen, e.g., as one of the  following:
\begin{eqnarray}
\label{eq:A}
{\mathbf A}^{uu}=\left(\begin{array}{*{2}{c}} 1 & 0\\ 0 & 0 \end{array} \right),\quad
{\mathbf A}^{uv}=\left(\begin{array}{*{2}{c}} 0 & 1\\ 0 & 0 \end{array} \right),\nonumber \\
{\mathbf A}^{vu}=\left(\begin{array}{*{2}{c}} 0 & 0\\ 1 & 0 \end{array} \right),\quad
{\mathbf A}^{vv}=\left(\begin{array}{*{2}{c}} 0 & 0\\ 0 & 1 \end{array} \right),
\end{eqnarray}
where the superscripts label the coupling scheme, e.g., ${\mathbf A}^{uu}$ denotes a matrix representing the coupling scheme $uu$.
The function $\ker\left(x\right)$ is an integral kernel satisfying:
\begin{eqnarray}
	\ker\left( x \right)=\ker\left( -x \right), \quad \int_{-\infty}^{\infty} \left| \ker\left(x\right) \right| dx = 1.\nonumber
\end{eqnarray}
The term
\begin{eqnarray}
	 \eta U\left(x\right) = \int \ker(x') dx' U\left(x\right) 
 \end{eqnarray}
in Eq.~\eqref{eq:control_term} is introduced in order to make the control noninvasive with respect to the homogeneous steady state, i.e., the homogeneous steady state is not changed by the control term $\hat K U$. In this paper we will focus on the following three symmetric integral kernels:
\begin{eqnarray}
	\ker\left(x\right) =& \frac{1}{2}\left(\delta\left(x+\sigma\right)+\delta\left(x-\sigma\right)\right),\label{eq:kerDelta}\\
	\ker\left(x\right) =& \frac{1}{2\sigma}e^{-\frac{\left| x \right|}{\sigma}},\label{eq:kerExp}\\
	\ker\left(x\right) =& \frac{N}{\sigma\sqrt{2\pi}}
		\left( \frac{1}{r}e^{-\frac{x^2}{2\left(r\sigma\right)^2}} - e^{-\frac{x^2}{2\sigma^2}}\label{eq:kerMexhat}
		\right), 0<r<1,
\end{eqnarray}
where $\sigma$ is a characteristic nonlocal coupling range, and $r=\sigma_e/\sigma_i<1$ describes the ratio of short-range excitatory ($\sigma_e$) and long-range inhibitory ($\sigma \equiv\sigma_i$) interaction in the Mexican hat kernel Eq.~(\ref{eq:kerMexhat}). The prefactor $N$ is chosen such that the kernel in Eq.~(\ref{eq:kerMexhat}) is normalized in the $L_1$ norm. For the
$\delta$-function Eq.~(\ref{eq:kerDelta}) and the exponential kernel Eq.~(\ref{eq:kerExp}) $\eta=1$ (to secure noninvasiveness), while for the Mexican hat kernel Eq.~(\ref{eq:kerMexhat}) $\eta=0$.
Figure~\ref{fig:kernels} shows plots of these kernels, and additionally an anisotropic $\delta$-function kernel 
$\ker\left(x\right) = \delta(x+\sigma)$ (corresponding to backward coupling only) and a rectangular kernel which simplifies to the symmetric $\delta$-function kernel in the limit of zero width $w$. Suppression of pulse propagation by isotropic (symmetric) and anisotropic (asymmetric) $\delta$-function integral kernels has already been studied by Schneider et al.\cite{SCH09c}. 

\section{Stability analysis}
Before investigating the effects of the control upon a traveling pulse, we will focus on the instabilities of the homogeneous steady state (HSS) first. This will allow us to get some analytical insight. After a survey of the possible spatio-temporal instabilities in excitable media we will analytically describe the stability borders in the control parameter space of the control strength $\kappa$ and the nonlocal coupling range $\sigma$.

\subsection{Instabilities of the homogeneous steady state}
\begin{figure}\resizebox{.5\textwidth}{!}{%
	\includegraphics[width=0.2\textwidth]{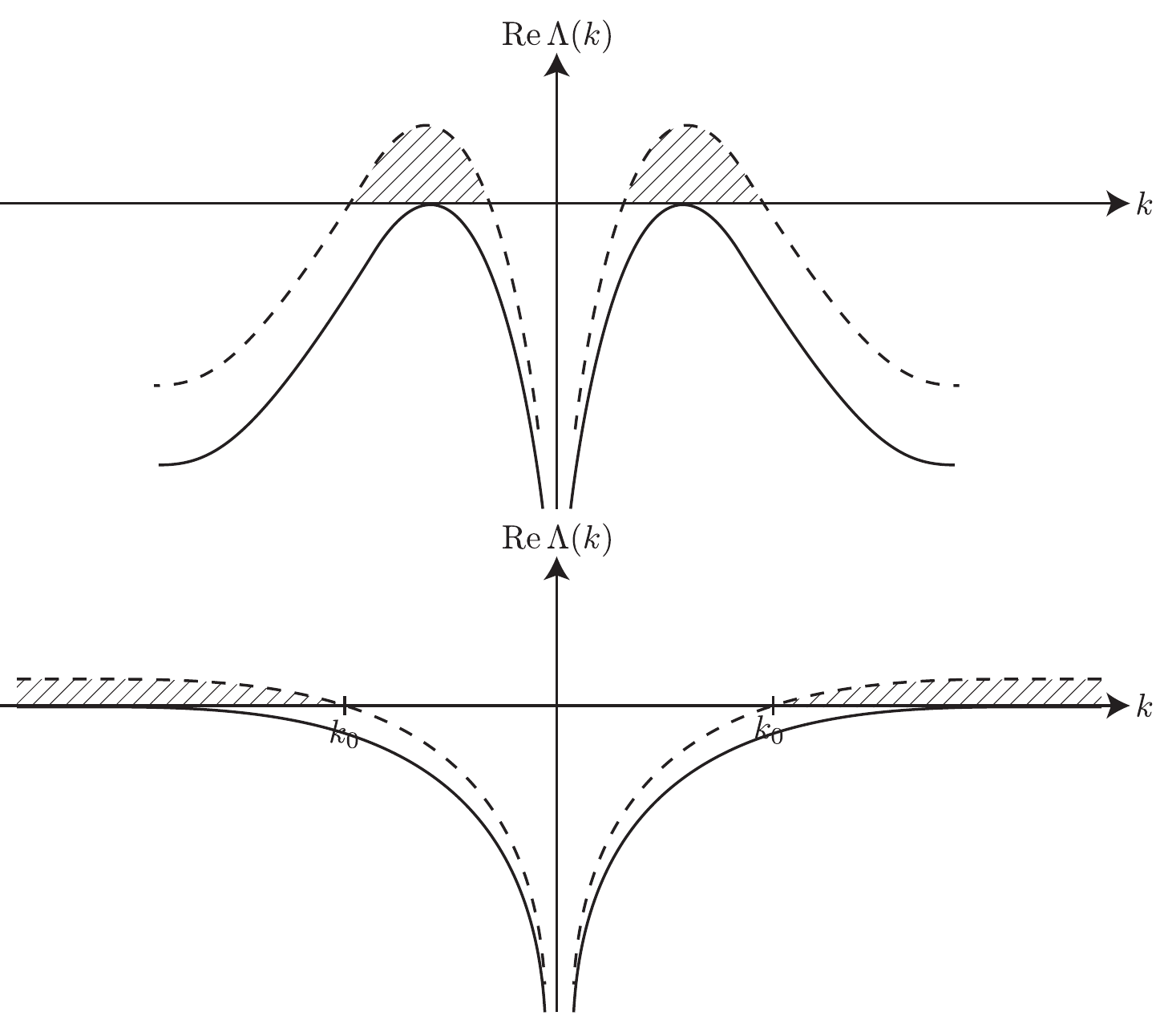}
}\caption{Schematic dispersion relation of the homogeneous steady state for Turing or wave instability (top) and salt-and-pepper instability (bottom). The solid and dashed lines mark $\Re\Lambda(k)$ at the bifurcation point and beyond the bifurcation, respectively. The hatched region corresponds to the band of unstable wavevectors beyond the bifurcation.}
\label{fig:instab}
\end{figure}

\begin{figure}
  \centering

  \begin{subfigure}[b]{0.3\textwidth}
    \includegraphics[width=\textwidth]{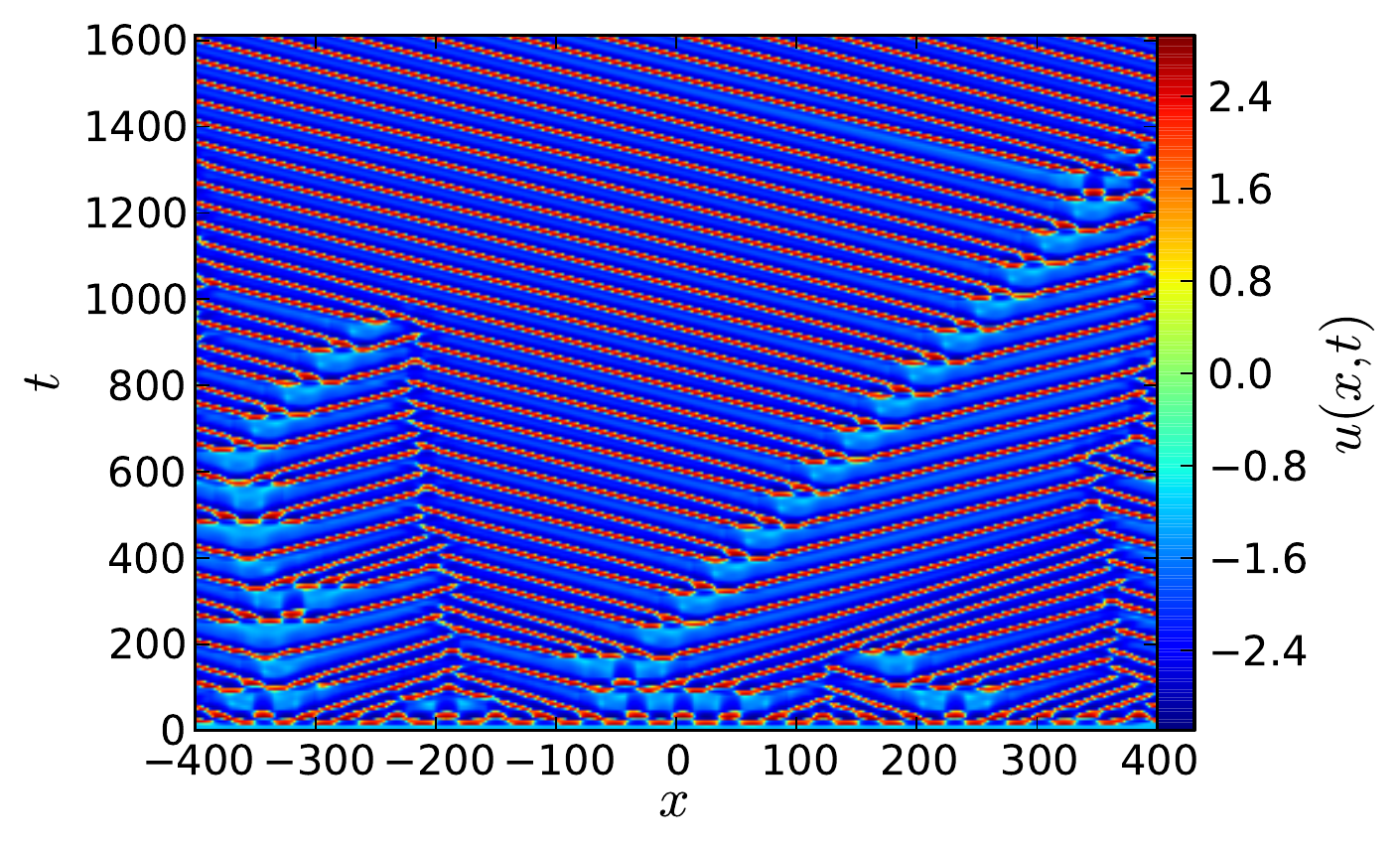}
	  \caption{(a)}
	  \label{fig:instab_wave}
  \end{subfigure}
  \begin{subfigure}[b]{0.3\textwidth}
	  \includegraphics[width=\textwidth]{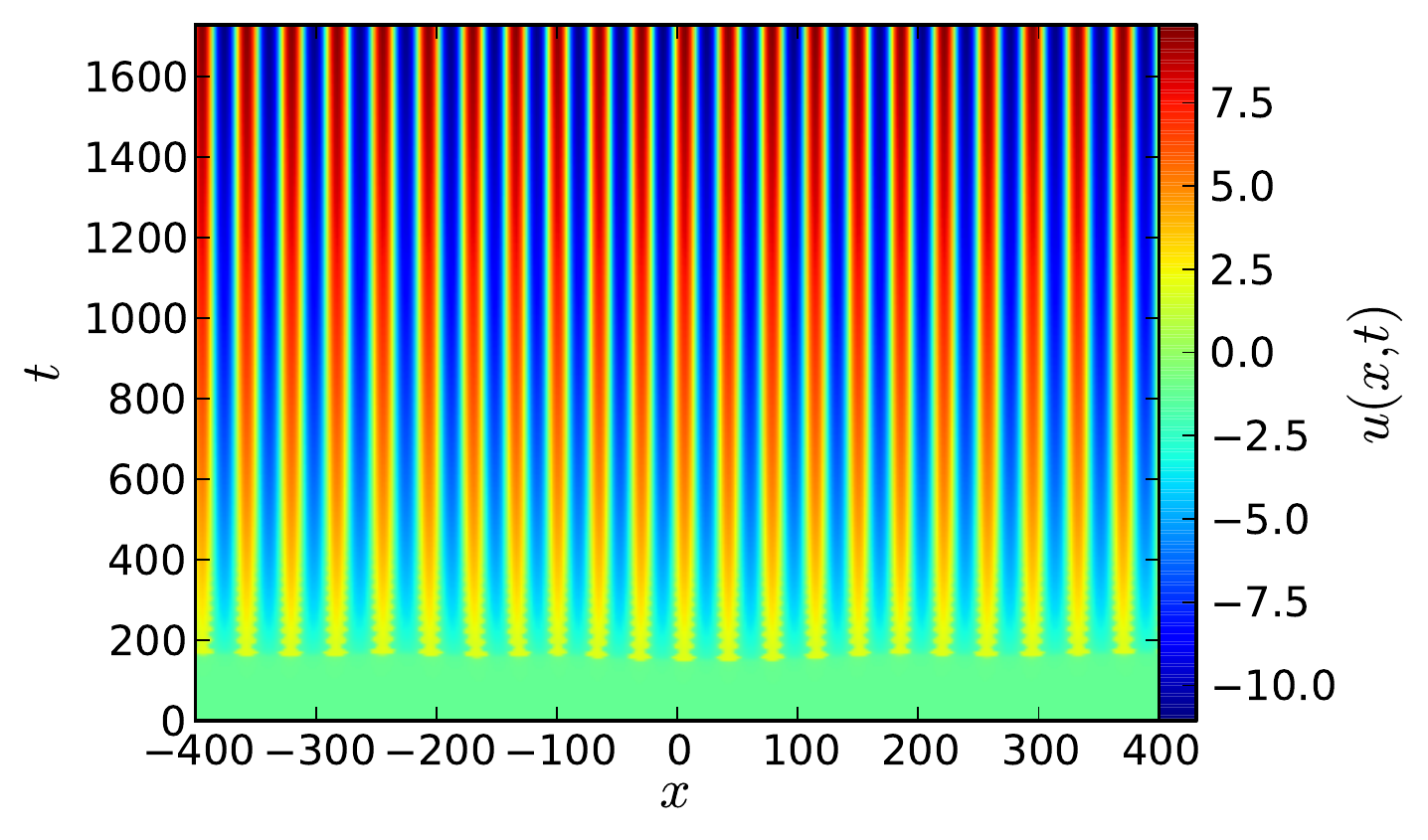}
		\caption{(b)}
	  \label{fig:instab_turing}
  \end{subfigure}
  \begin{subfigure}[b]{0.3\textwidth}
	  \includegraphics[width=\textwidth]{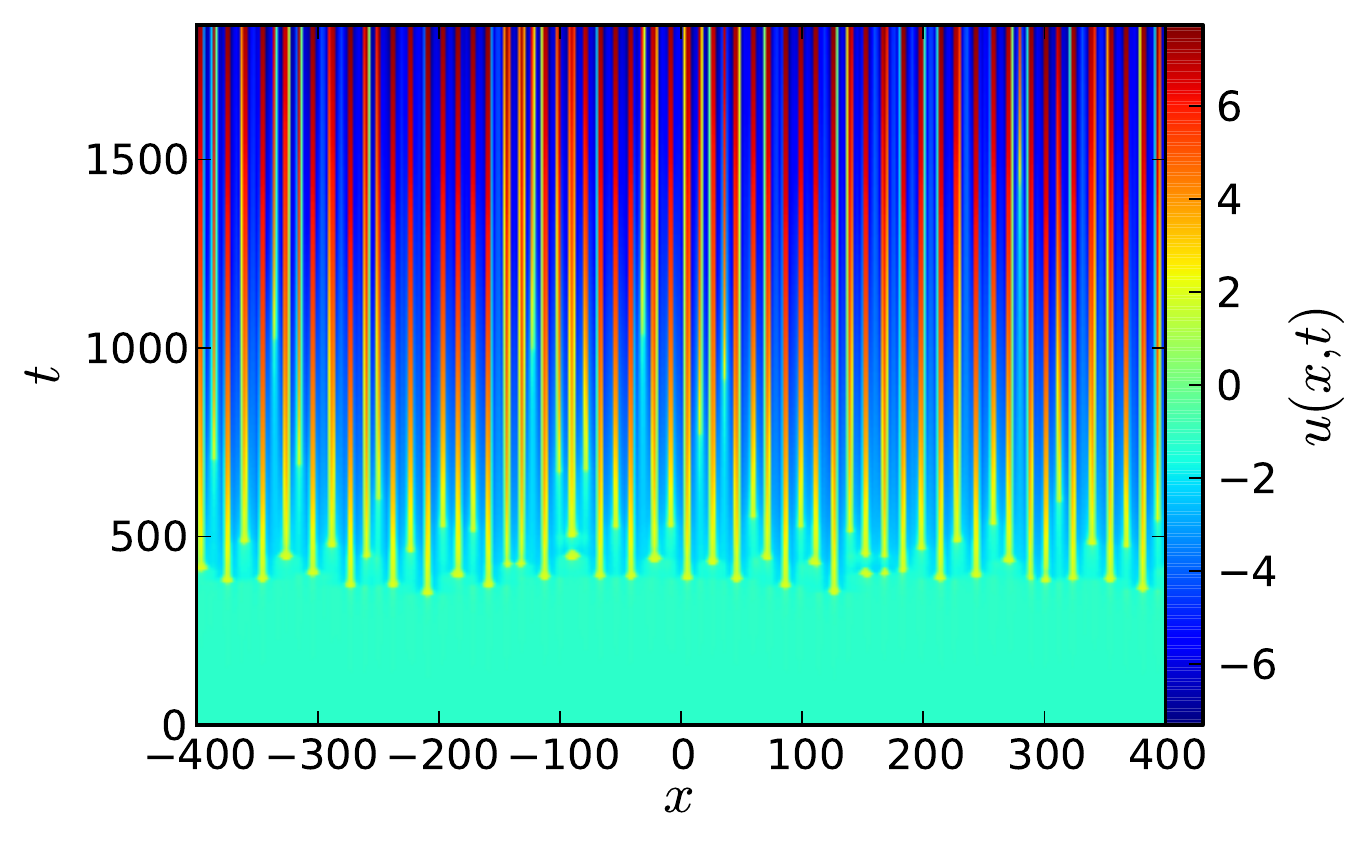}
		\caption{(c)}
	  \label{fig:instab_degenerate}
  \end{subfigure}
\caption{Space-time plots of the onset of spatio-temporal instabilities of the homogeneous steady state caused by nonlocal control. As initial condition weak white noise is added to the homogeneous steady state. The control is switched on at $t=0$. (a) Wave instability: Mexican hat kernel, $uu$-coupling-scheme, $r=\sigma_{e}/\sigma_{i}=0.9$, $\left(\kappa,\sigma_{i}\right)=\left(1.5,\Delta x\right)$, (b) Turing instability: Mexican hat kernel, $uv$-coupling-scheme, $r=\sigma_{e}/\sigma_{i}=0.9$, $\left(\kappa,\sigma_{i}\right)=\left(1.75,\Delta x\right)$, (c) Salt-and-pepper instability: Exponential kernel, $uv$-coupling-scheme, $\left(\kappa,\sigma\right)=\left(-1.25,2\Delta x\right)$. Parameters: 
$\varepsilon=0.08$, $\beta= 1.2$, $L=800$, simulation timestep $dt=0.005$.}
\label{fig:instabilities_sim}
\end{figure}

\begin{figure}
\resizebox{.5\textwidth}{!}{%
\includegraphics[width=0.2\textwidth]{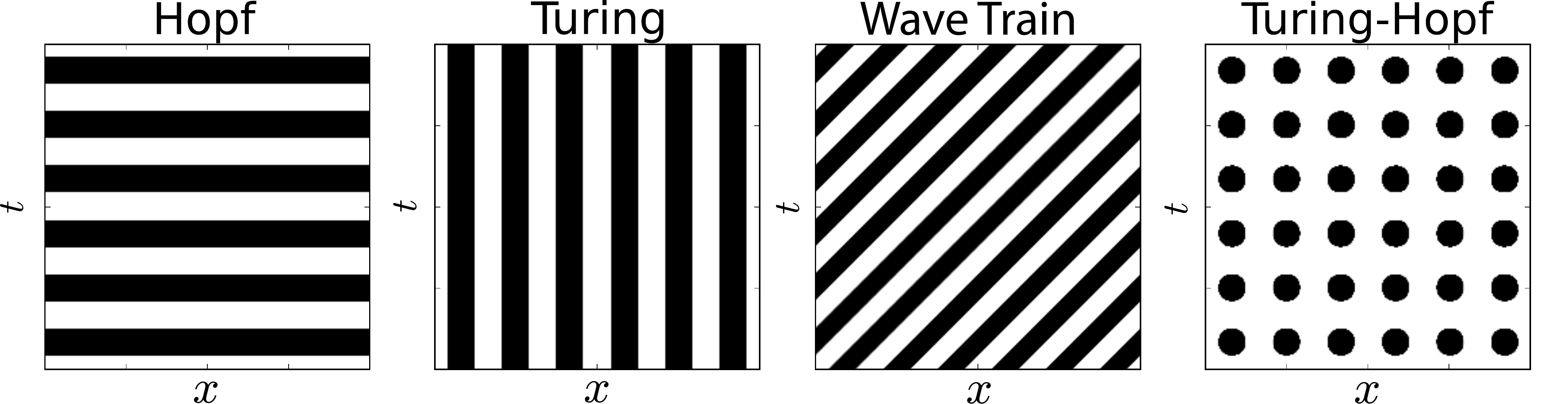}
}
\caption{Schematic space-time patterns in reaction-diffusion systems corresponding to the instabilities of the homogeneous steady state: Hopf, Turing, wave train, Turing-Hopf (from left to right).}
\label{fig:spatiotemporal_schematic}
\end{figure}

Generally, the stable homogeneous steady state of a spatially extended system can become unstable by different spatio-temporal bifurcations
when the control parameters are changed. We characterize the different instabilities by the behavior of the dispersion relation obtained from a linear stability analysis of the homogeneous steady state, i.e., the eigenvalue $\Lambda\left(k\right)$ in dependence upon the wavenumber $k$. Depending upon the value $k_c$ at which the bifurcation ($\Re\Lambda\left(k_c\right)=0$) occurs and the corresponding imaginary part of the eigenvalue $\Im\Lambda\left(k_c\right)$, four different cases can be distinguished: (i) A Hopf instability causing spatially homogeneous oscillations occurs for $k_c=0$ and $\Im\Lambda(k_c)\neq 0$. (ii) A Turing instability is given when $k_c\neq0$ and $\Im\Lambda(k_c)=0$, which leads to a stationary spatial modulation. (iii) A wave instability is characterized by $k_c\neq0$ and $\Im\Lambda(k_c)\neq0$. Both (ii) and (iii) have in common that upon further increase of the bifurcation parameter there appears a finite interval of unstable wavenumbers $[k_-,k_+]$:
\begin{eqnarray}
	\Re\Lambda\left(k\right)>0\quad\forall \left|k\right|\in [k_-,k_+].,\label{eq:k_finite}
\end{eqnarray}
(iv) If the instability $\Re\Lambda\left(k_c\right)=0$ occurs at $k_c \to \infty$, 
the condition (\ref{eq:k_finite}) is not fullfilled. Rather, the following degenerate case holds upon further increase of the bifurcation parameter beyond the bifurcation:
\begin{eqnarray}
	\Re\Lambda\left(k\right)>0\quad\forall \left|k\right| >k_0,\label{eq:k0}
\end{eqnarray}
Thus arbitrarily large wavenumbers become unstable, which can cause patterns with arbitrarily short wavelengths. Such instabilities
have been called {\em salt-and-pepper patterns} \cite{KON10}, and it has been suggested that they might occur in morphogenesis when differentiated cells inhibit the differentiation of neighboring cells, as is seen, for example, with differentiated neuroprogenitor cells in the epithelium of Drosophila embryos \cite{KON10}.
Figure~\ref{fig:instab} illustrates the difference of the Turing or wave instability (upper panel) and the salt-and-pepper instability (lower panel). A tabular overview of the instabilities (i)-(iv) is given in Table~\ref{tab:instabilities}.
Three examples of the spatio-temporal patterns generated by the instabilities (ii)-(iv) are provided in
Figure~\ref{fig:instabilities_sim} from simulations of the full nonlinear equations (\ref{eq:main}). After some initial transients
a wave instability (a), a Turing instability (b), and a salt-and-pepper instability (c) develops. In each case the initial condition
at $t=0$ is the homogeneous steady state to which Gaussian white noise is added. 
Fig.~\ref{fig:spatiotemporal_schematic} shows schematically the space-time patterns corresponding to the instabilities (i) - (iii) of the homogeneous steady state in reaction-diffusion systems: Hopf, Turing, wave train; additionally, the pattern on the very right corresponds to a combined Turing-Hopf codimension-two instability \cite{MEI97a}.

\begin{table}
	\caption{Instabilities of the homogeneous steady state that give rise to pattern formation in reaction-diffusion systems}
	\label{tab:instabilities}       
	\begin{tabular}{lll}
		\hline\noalign{\smallskip}
		Instability & $\left|k_c\right|$ & $\Im\Lambda\left(k_c\right)$\\
		\noalign{\smallskip}\hline\noalign{\smallskip}
		Hopf & $0$ & $\neq 0$ \\
		Turing & $\neq 0$, finite & $=0$ \\
		Wave & $\neq 0$, finite & $\neq0$ \\
		Salt-and-pepper & infinite;&\\& $\exists k_0: \Re\Lambda^+(k)>0 \quad \forall \left| k \right| > k_0$ & - -\\
		\noalign{\smallskip}\hline
	\end{tabular}
\end{table}

\subsection{Stability boundaries in control parameter space}
We can obtain an analytical expression for the stability boundaries of the homogeneous steady state in control parameter space $\left(\kappa,\sigma\right)$ with the help of a linear stability analysis. We linearize the system~\eqref{eq:main} about the homogeneous steady state $U^{*}_\text{hom}$ for small perturbations $\delta U$. The resulting linear differential equation can be solved with the ansatz  $\delta U \sim e^{\Lambda t}e^{i k x}$, which yields the following characteristic equation:
\begin{eqnarray}
	0=\det\left[
	\left(\begin{array}{cc}
\Lambda+b\left(k\right) & 1\\
-1 & \varepsilon^{-1}\Lambda
\end{array}\right)\right. \nonumber\\
	\left.-\kappa A\left(\tilde{g}\left(k\right)-\eta\right)
		\right],\label{eq:char}
	\end{eqnarray}
where $b(k)=k^{2}+\beta^2-1>0$ and $\tilde{g}$ is the Fourier transform of the integral kernel:
\begin{eqnarray}
	\tilde{g}(k) = \int_{-\infty}^{\infty} \ker\left(x\right)e^{-ikx}dx.
\end{eqnarray}
For further analytical progress we need to specify the coupling scheme. Here we will present the calculations for the activator self-coupling scheme ${\mathbf A}^{uu}$ and the exponential integral kernel. Similar results have been obtained for the other coupling schemes and integral kernels ~\cite{BAC13} . The characteristic equation~\eqref{eq:char} reads for the $uu$-coupling scheme:
\begin{eqnarray}
	\Lambda^{2}+\Lambda\left(b\left(k\right)-\kappa\left(\tilde{g}(k)-\eta\right)\right)+\varepsilon&=&0
\end{eqnarray}
which yields the dispersion relation
\begin{eqnarray}
	\Lambda_{\pm}\left(k\right) = \frac{1}{2} \left\{
		-b\left(k\right) + \kappa\left(\tilde{g}(k)-\eta\right) \right.\nonumber\\
		\left.\pm
		\sqrt{\left[b\left(k \right) - \kappa\left(\tilde{g}(k)-s\right)\right]^{2} - 4\varepsilon}
		\right\}.
\end{eqnarray}
For a given control configuration we determine the stability of the homogeneous steady state
by calculating the maximum real part of $\Lambda_+\left(k\right)$, as $\Re\left(\Lambda_+\right)>\Re\left(\Lambda_-\right)$. 
Let us assume that a control configuration consisting of the coupling scheme $A$, the integral kernel $\ker (x)$ with the coupling range $\sigma_0$, and the coupling strength $\kappa_0$ are given. This configuration destabilizes the homogeneous steady state if
\begin{eqnarray}
	\exists k\in\mathbb{R}:\mu=\max\Re\left(\Lambda_\pm\left(k\right)\right)>0.\label{eq:instability_condition}
\end{eqnarray}
\begin{figure}
	\resizebox{0.4\textwidth}{!}{%
    \includegraphics[width=0.1\textwidth]{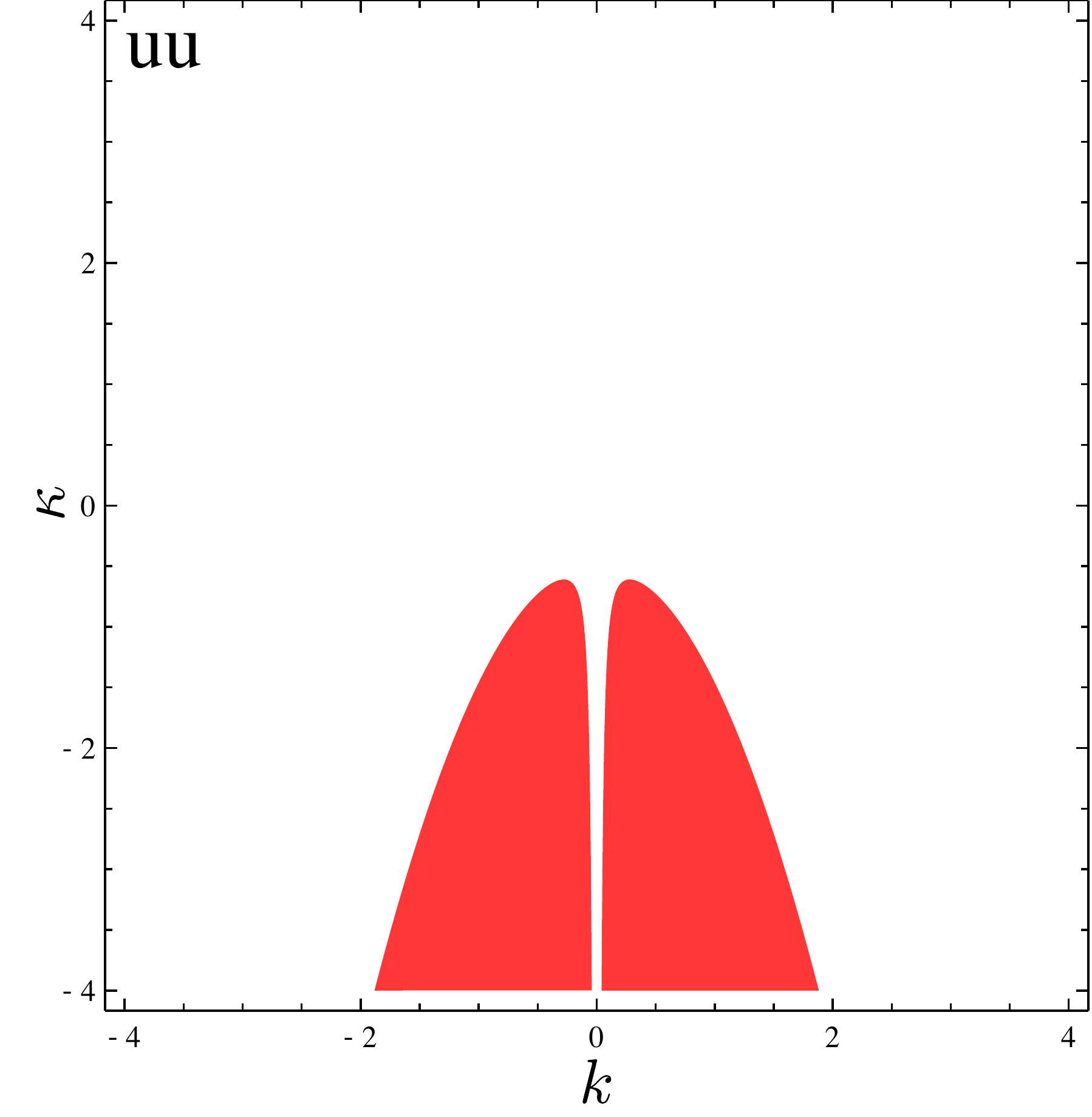}	
	}
	\caption{Stability regimes of the homogeneous steady state. Red indicates control parameters $(\kappa,\sigma)$ destabilizing the homogeneous steady state. Control parameters: $\sigma=\Delta x$, exponential kernel, ${uu}$ coupling scheme. System parameters: $\varepsilon=0.08, \beta=1.2$.}
	\label{fig:exp_unstable}
\end{figure}

In Fig.~\ref{fig:exp_unstable}, the stability of the homogeneous steady state is calculated numerically for the exponential kernel in the $uu$-coupling scheme for $\sigma=\Delta x$. The control strength $\kappa$ is varied and so is the wavenumber $k$. Regimes with an unstable homogeneous steady state are colored red. There is a critical control strength $\kappa_c(\sigma)=-0.61$.  For $\kappa<\kappa_c$  the control destabilizes the homogeneous steady state, whereas the stability of the homogeneous steady state is not affected for $\kappa>\kappa_c$. The critical wave number $k_c$ is the key to the analytical description of the stability border in control parameter space $\left(\kappa,\sigma\right)$. The stability border is described by the set of  control parameters $\left(\kappa,\sigma\right)$ for which we can find a $k$ fulfilling:
\begin{eqnarray}
	\Re\left( \Lambda_+\left(k\right)\right) &=0,\label{eq:cond1}\\
	\partial_k\Re\left( \Lambda_+\left(k\right)\right)&=0.\label{eq:cond2}
\end{eqnarray}
Using Eq.~\eqref{eq:cond1} we obtain a closed analytical expression:
\begin{eqnarray}
	\kappa=\frac{b\left(k\right)}{\tilde{g}\left(k\right)-\eta}.\label{eq:1}
\end{eqnarray}
Substitution into Eq.~\eqref{eq:cond2} gives:
\begin{eqnarray}
	-2k + b\left(k\right)\frac{\partial_k\tilde{g}\left(k\right)}{\tilde{g}\left(k\right)-\eta}=0.\label{eq:2}
\end{eqnarray}
Using Eq.~\eqref{eq:2} and substituting $\tilde{g}\left(k\right)=\left(1+k^2\sigma^2\right)^{-1}$ and $\eta=1$ for the exponential kernel we find:
\begin{eqnarray}
	k_c= \sqrt[4]{\frac{\beta^2-1}{\sigma^2}},
\end{eqnarray}
which gives us the stability border together with Eq.~\eqref{eq:1}:
\begin{eqnarray}
	\sigma\left(\kappa\right)=\frac{1}{\sqrt{-\kappa}-\sqrt{\beta^2-1}},
\end{eqnarray}
where $\kappa<1-\beta^2$ as $k_c\in\mathbb{R}$:
\begin{eqnarray}
	k_c=\pm\sqrt[4]{a}\sqrt{\sqrt{-\kappa}-\sqrt{\beta^2-1}}.
\end{eqnarray}
Thus, we have found an analytical expression for the stability boundary in control parameter space $(\kappa,\sigma)$ . As the critical wavenumber $k_c\neq0$, the emerging instability is either a Turing, a wave, or a salt-and-pepper instability. The latter can be excluded because of the following asymptotic behavior for large $k$:
\begin{eqnarray}
	\Re\Lambda_+\left(k\gg1\right) = \Lambda_+\left(k\right) = -\frac{\varepsilon}{k^2}+\mathcal{O}\left(k^{-4}\right)<0.
\end{eqnarray}
This result implies that we cannot find a $k_0$ fulfilling Eq.~\eqref{eq:k0} and, thus, we can exclude the salt-and-pepper instability
for this type of kernel and coupling scheme.

\section{Simulation of pulses}
In this section we summarize our results of the simulation of Eq.~\eqref{eq:main} for a wide range of parameters 
(strength $\kappa$ and range $\sigma$ of the nonlocal coupling) for the four different coupling schemes Eqs.~(\ref{eq:A}) with the three control kernels presented in Eqs.~(\ref{eq:kerDelta})-(\ref{eq:kerMexhat}).
Various types of space-time patterns can arise if a pulse in the uncontrolled system, i.e., without nonlocal coupling, is used as initial condition, and then the nonlocal coupling is switched on. 

\begin{figure}
	\resizebox{0.5\textwidth}{!}{
		\includegraphics[width=0.5\textwidth]{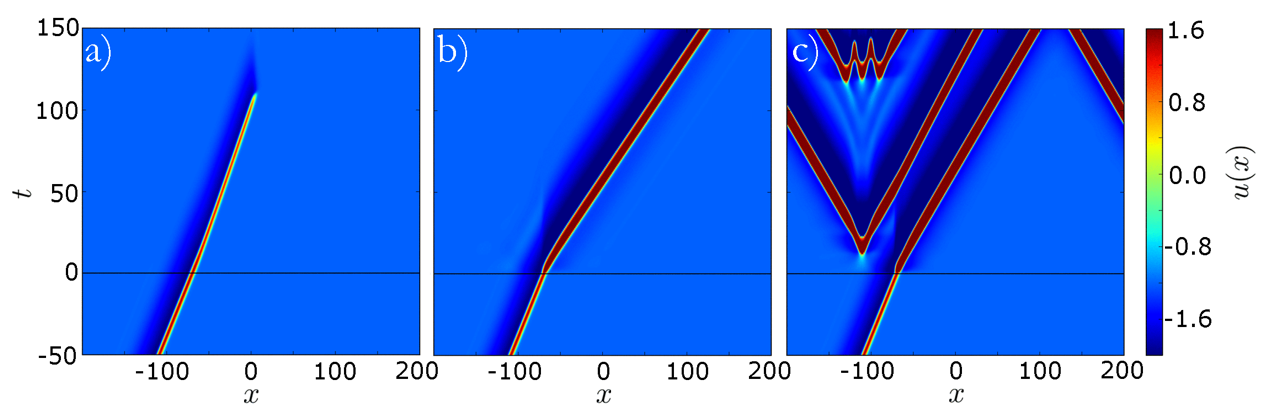}
}
\caption{Examples of the effects of the nonlocal control on a traveling pulse: space-time plots for (a) pulse suppression, (b) pulse acceleration, (c) multiple pulse generation. The control is switched on at $t=0$. 
(a) Exponential kernel, $vv$-coupling-scheme, $\left(\kappa,\sigma_{i}\right)=\left(0.5, 0275\Delta x\right)$, (b) Mexican hat kernel, $uu$-coupling-scheme, $r=\sigma_{e}/\sigma_{i}=0.1$, $\left(\kappa,\sigma_{i}\right)=\left(0.5, 2\Delta x\right)$, (c) Exponential kernel, $uu$-coupling-scheme, $\left(\kappa,\sigma_{i}\right)=\left(-0.75, 1.5\Delta x\right)$. Other parameters: $\left(\varepsilon,\beta\right)=\left(0.08,1.20\right)$.}
	\label{fig:control_effects}
\end{figure}

The initial condition is the stable traveling pulse that forms after an excitation of the uncontrolled system. In all simulations the boundary conditions are periodic. 
The diffusion is implemented with the spectral method \cite{CRA06}. At time $t=0$ the control is switched on. For every parameter configuration, the effect of the control term may be classified as one of the following:
\begin{description}
 \item[(a)] Pulse suppression (PS),
 \item[(b)] Pulse acceleration or deceleration,
 \item[(c)] Multiple pulse generation (MP).
\end{description}
Examples of space-time plots for each class are provided in Fig.~\ref{fig:control_effects}. Panel (a) shows pulse suppression, (b) shows pulse acceleration, and (c) shows the generation of a pair of pulses propagating into opposite directions. Figs.~\ref{fig:iso_delta}-\ref{fig:mexhat} summarize the results for the different regimes in the $\left(\kappa,\sigma\right)$ control parameter plane for the four coupling schemes and the $\delta$-function kernel (Fig.~\ref{fig:iso_delta}), the exponential kernel (Fig.~\ref{fig:exp}), and the Mexican hat kernel (Fig.~\ref{fig:mexhat}). In addition, the analytically calculated stability boundary of the homogeneous steady state is plotted as a white line, and the corresponding instabilities are indicated by white hatched areas: wave instability (diagonal hatching), Turing instability (vertical hatching), salt-and-pepper instability (dotted hatching). For exemplary space-time plots corresponding to these instabilities, refer to Fig.~\ref{fig:instabilities_sim}(a),(b),(c), respectively.

\subsection{Isotropic $\delta$-function kernel}
\begin{figure}
	\resizebox{0.5\textwidth}{!}{
		\includegraphics{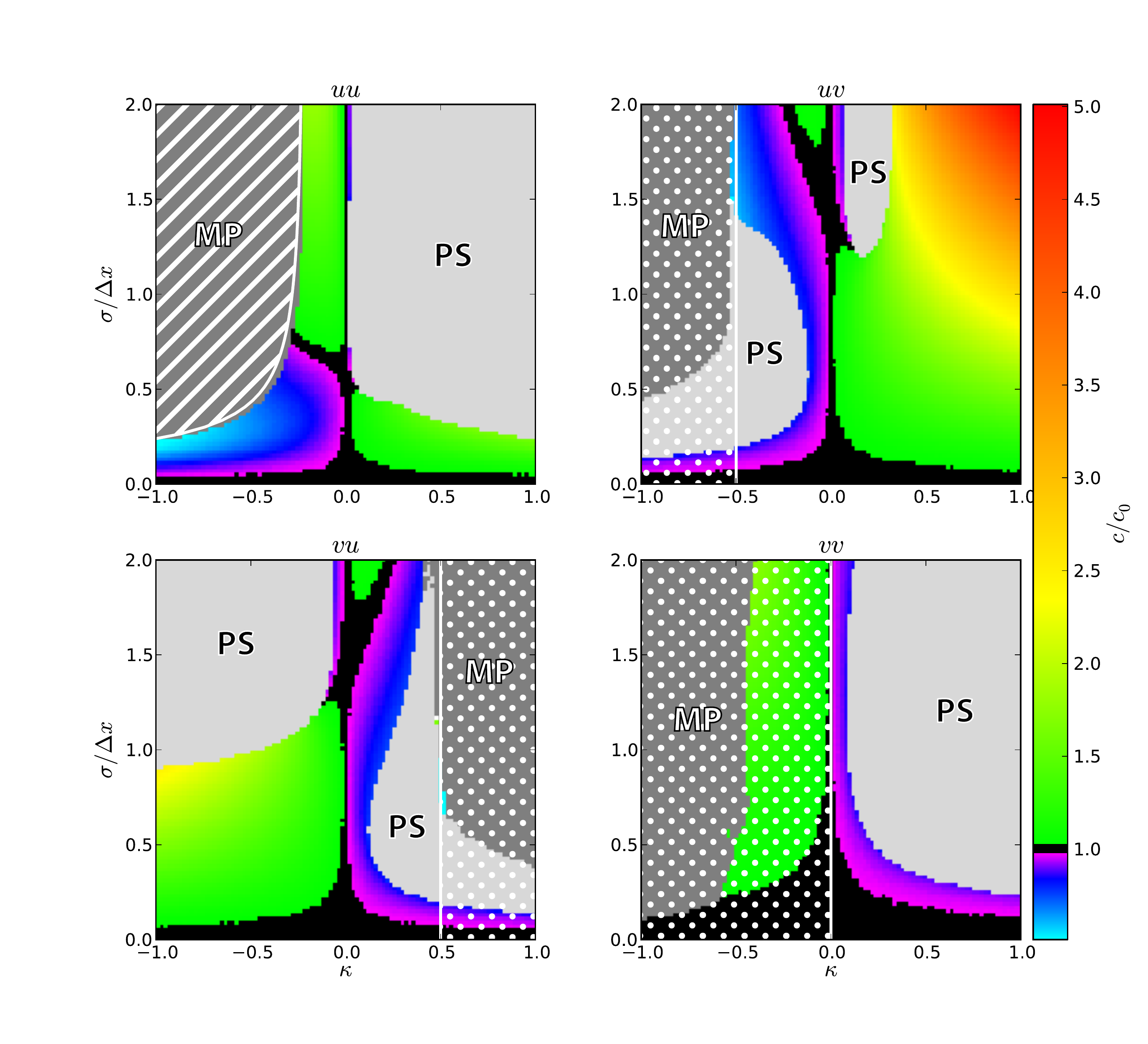}}
	\caption{Effects of the nonlocal coupling upon pulse pulse propagation. The four panels summarize the resulting space-time patterns in the control parameter plane of the coupling strength $\kappa$ and range $\sigma$ for the four coupling schemes ($uu$, $uv$, $vu$, $vv$) with the $\delta$-function integral kernel Eq.(\ref{eq:kerDelta}). Dark gray: Multiple pulse generation (MP). Light gray: Pulse suppression (PS). Colors blue to red: Pulse acceleration or deceleration (the pulse velocity $c$ normalized by the uncontrolled velocity $c_0$ is color coded). Black: Pulse speed changes less than $0.5\permil$ compared to the uncontrolled pulse. The regions with white hatching indicate an unstable homogeneous steady state due to one of the following instabilities: wave instability (diagonal hatching), 
salt-and-pepper instability (dotted hatching). System parameters: $\left(\varepsilon,\beta\right)=\left(0.08,1.2\right)$. $L=600$, integration timestep $dt=0.005$, spatial resolution $dx=L/16384\approx0.04$.}
\label{fig:iso_delta}
\end{figure}
We will now discuss the effects of nonlocal control systematically and in detail. First, we consider nonlocal coupling with the $\delta$-function integral kernel Eq.(\ref{eq:kerDelta}).
In the $uu$-coupling scheme (Fig.~\ref{fig:iso_delta}, top left), pulse suppression (PS), cf. Fig.~\ref{fig:control_effects}(a), can only be achieved with positive control strength $\kappa$ (light gray region in the ($\kappa, \sigma$) plane). This is in accordance with \cite{SCH09c}. For negative control strength there is a large regime of multiple pulse (MP) generation (dark gray region in the ($\kappa, \sigma$) plane), cf. Fig.~\ref{fig:control_effects}(c). The border to this regime is well described by the analytically calculated stability boundary of the homogeneous steady state. The corresponding instability (diagonal hatching) is a wave instability, see the exemplary space-time plot in Fig.~\ref{fig:instabilities_sim}(a), and it coincides with the numerically observed MP regime. The other regions in the ($\kappa, \sigma$) control parameter space correspond to propagation of the initially existing pulse with a changed velocity, where the pulse velocity is color coded. For small $\sigma$ and positive $\kappa$ the pulse is accelerated (green), cf. Fig.~\ref{fig:control_effects}(b), while
for small $\sigma$ and negative $\kappa$ the pulse is slowed down (violet and blue). This behavior changes when $\sigma$ is increased, i.e., pulse suppression is found for positive $\kappa$, while pulse acceleration or multiple pulse generation occurs for negative $\kappa$. Only in a small part of parameter space (small $|\kappa|$ or small $\sigma$) the pulse velocity remains unchanged (black). 

The parameter planes for the $uv$- (top right) and $vu$-coupling schemes (bottom left) are related to each other by an approximate reflection symmetry with respect to $\kappa \to -\kappa$, at least for not too large $\sigma$, and thus they show qualitatively similar behavior. In contrast to $uu$-coupling, pulse suppression can be achieved for both negative and positive control strengths. In case of $uv$- ($vu$-)coupling, the stability boundary of the homogeneous steady state is described by the vertical straight line $\kappa(\sigma)=-0.5$ ($\kappa(\sigma)=0.5$, respectively). The homogeneous steady state becomes unstable due to a salt-and-pepper instability (dotted hatching) to the left or right of that line, respectively; see the exemplary space-time plot in Fig.~\ref{fig:instabilities_sim}(c). One of the regimes of pulse suppression is partially covered by the region where the homogeneous steady state is unstable, and so is the regime of multiple pulse generation, indicating that nonlocal feedback control can suppress pulses or generate multiple pulses while destabilizing the homogeneous steady state. Note, however, that pulse suppression is also possible if the homogeneous steady state is stable, as in the $uu$-coupling scheme. Pulse acceleration (green) and deceleration (violet and blue) for small coupling range $\sigma$ is similar to the $uu$-coupling scheme in case of $uv$, and inverted with respect to $\kappa$ in the case of $vu$.
 
In the $vv$-coupling scheme pulse suppression is possible only for positive $\kappa$. Control configurations with negative control strength $\kappa$ destabilize the homogeneous steady state by a salt-and-pepper instability, where multiple pulse generation or pulse acceleration (green) are possible. 

\subsection{Exponential kernel}
\begin{figure}
	\resizebox{0.5\textwidth}{!}{
	  \includegraphics{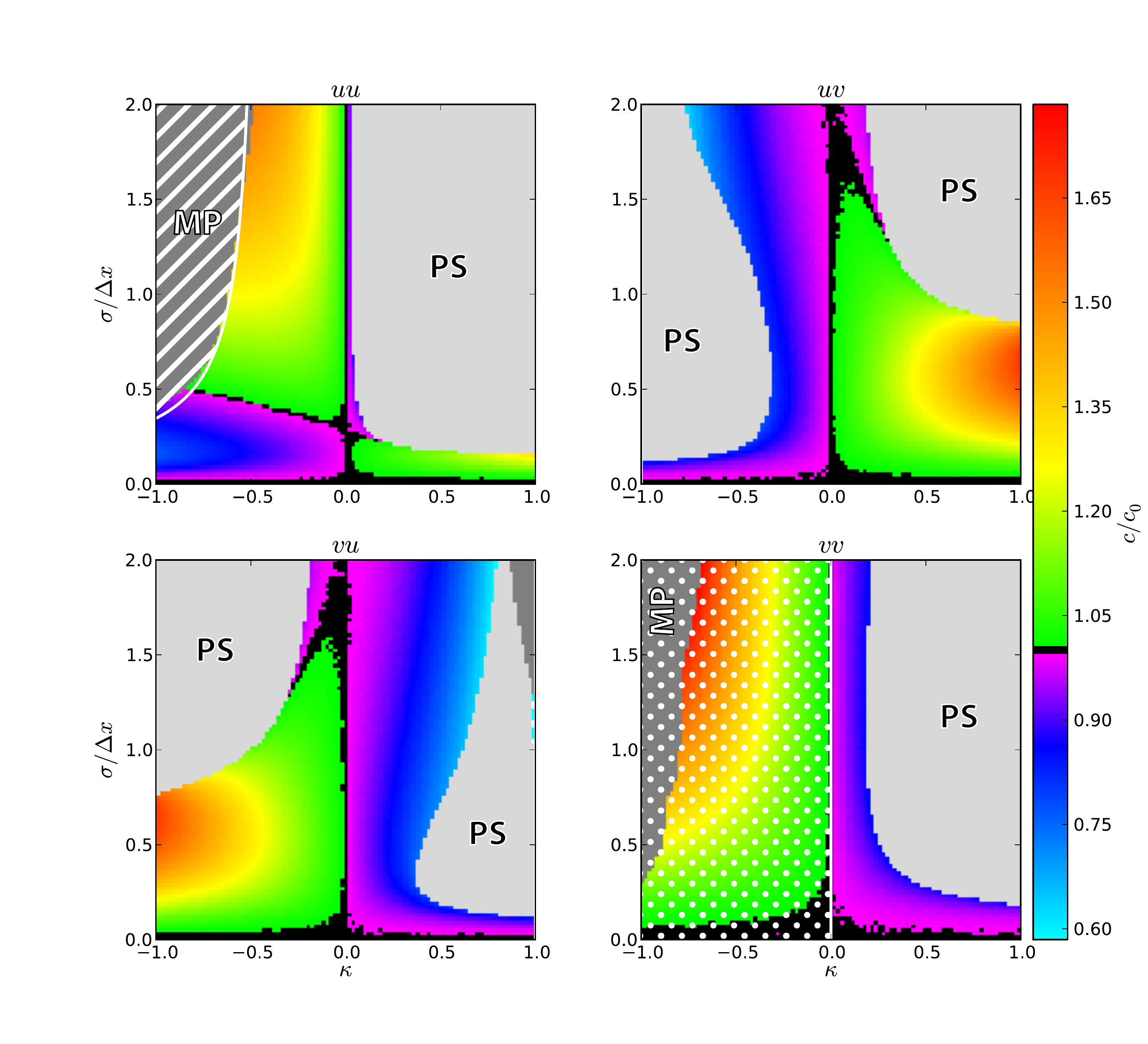}}
	\caption{Same as Fig.~\ref{fig:iso_delta} for the exponential integral kernel Eq.(\ref{eq:kerExp}).}
	\label{fig:exp}
\end{figure}
Next, we consider nonlocal coupling with the exponential integral kernel Eq.(\ref{eq:kerExp}) (Fig.~\ref{fig:exp}). The regimes in the control parameter plane for the exponential kernel function  are qualitatively similar to the ones of the isotropic $\delta$-function kernel. Pulse acceleration or deceleration and pulse suppression are found for the same signs of control strength. In the $uu$-coupling scheme the border to the regime of multiple pulse generation is, again, well described by the analytically calculated stability boundary of the homogeneous steady state (wave instability, white hatching). For the $uv$- and $vu$-coupling scheme the analytical stability boundary of the homogeneous steady state is given by $\kappa\left(\sigma\right)=-1$ and $\kappa\left(\sigma\right)=1$, respectively, and a salt-and-pepper instability arises for $\kappa < -1$ and $\kappa > 1$, respectively (outside the range plotted). There are no multiple pulses in the parameter range shown in the figure. The reflection symmetry of the control parameter planes of the $uv$- and $vu$-coupling schemes with respect to $\kappa \to -\kappa$ is much more pronounced, as compared to the $\delta$-function kernel. In the $vv$-coupling scheme the homogeneous steady state is unstable for $\kappa<0$ due to a salt-and-pepper instability, and the pulse acceleration can become much stronger than for the $\delta$-function kernel. 

\subsection{Mexican hat kernel}
\begin{figure}
	\resizebox{0.5\textwidth}{!}{
		\includegraphics{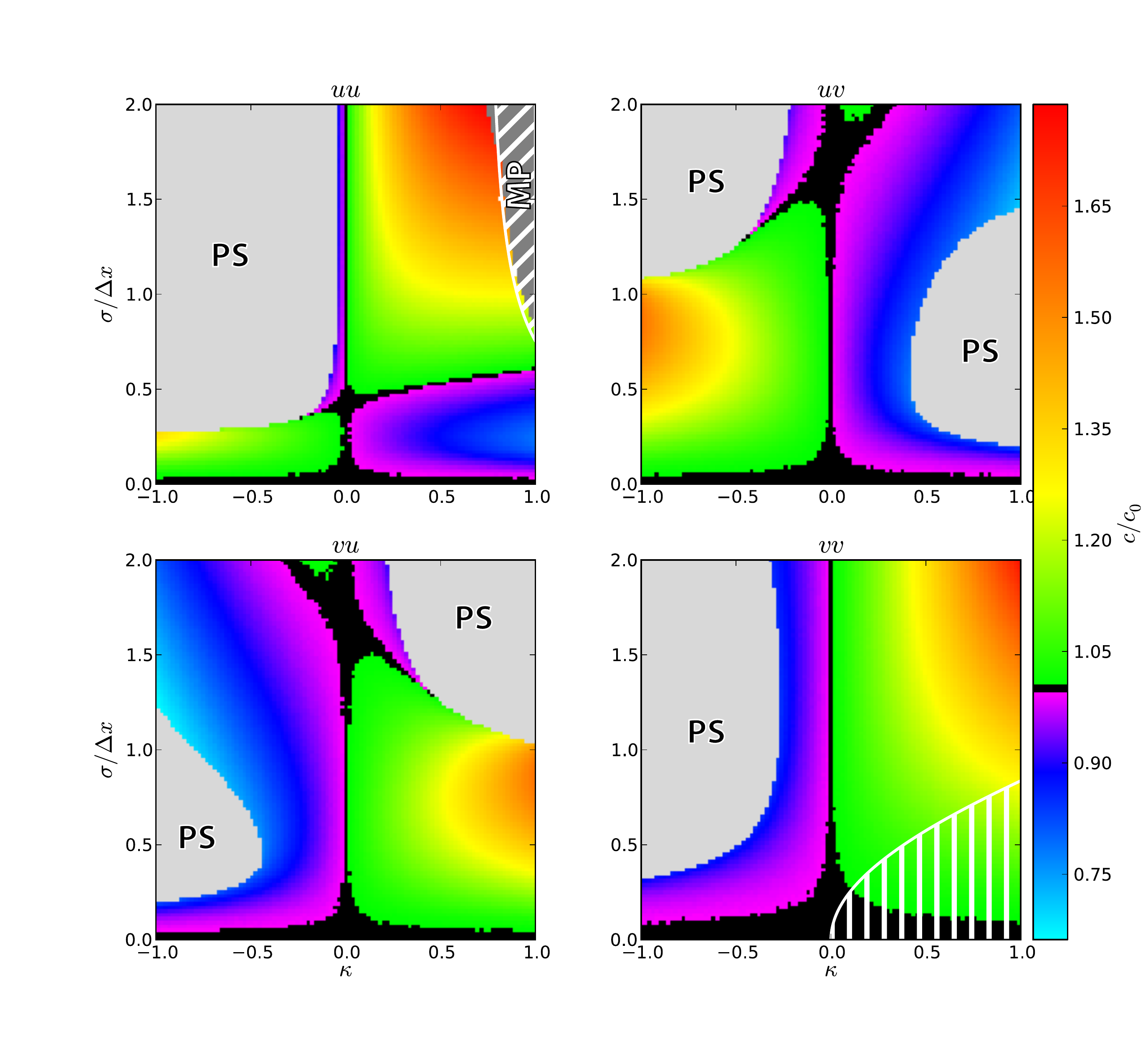}	}
	\caption{Same as Fig.~\ref{fig:iso_delta} for the Mexican hat integral kernel Eq.(\ref{eq:kerMexhat}) with $r=\sigma_{e}/\sigma_{i}=0.1$. White vertical or diagonal hatching indicates an unstable homogeneous steady state due to a Turing or a wave instability, respectively.}
	\label{fig:mexhat}
\end{figure}
Finally, we consider nonlocal coupling with the Mexican hat integral kernel Eq.(\ref{eq:kerMexhat}) (Fig.~\ref{fig:mexhat}). The Mexican hat kernel differs from the other kernels, since it is not positive definite. First of all, the term $\eta U\left(x\right)$ vanishes, as $\eta=0$. Second, the second moment of the kernel
\begin{eqnarray}
	M_2=\frac{1}{2}\int^\infty_{-\infty}y^2\ker(y)dy\nonumber
\end{eqnarray}
is negative while it is positive for the other kernel functions. This is the reason why the regimes in the control parameter plane are qualitatively similar as for the previous kernels if $\kappa$ is replaced by $-\kappa$: The regimes of pulse suppression are found for negative coupling strength $\kappa$ in the self-coupling schemes $uu$ and $vv$, and on the opposite sides of $\kappa$ in the cross-coupling schemes ($uv, vu$), compared to the other kernels, and the same holds for acceleration and deceleration of pulses. It should be noted that the instabilities of the homogeneous steady state in the $vv$-coupling scheme are of Turing type 
(see the exemplary space-time plot in Fig.~\ref{fig:instabilities_sim}(b)) for the Mexican hat kernel, whereas they are of the salt-and-pepper type for the other kernels. Turing instabilities also arise in the $uv$-coupling scheme for $\kappa\left(\sigma\right)>1.689$ ($r=0.1$), $>1.366$ ($r=0.5$), $>1.317$ ($r=0.9$), and in the $vu$-coupling scheme for negative $\kappa\left(\sigma\right)$ smaller than the corresponding negative threshold values (not shown in Fig.~\ref{fig:mexhat} in the range plotted).

\begin{figure}
	\resizebox{0.5\textwidth}{!}{
	  \includegraphics{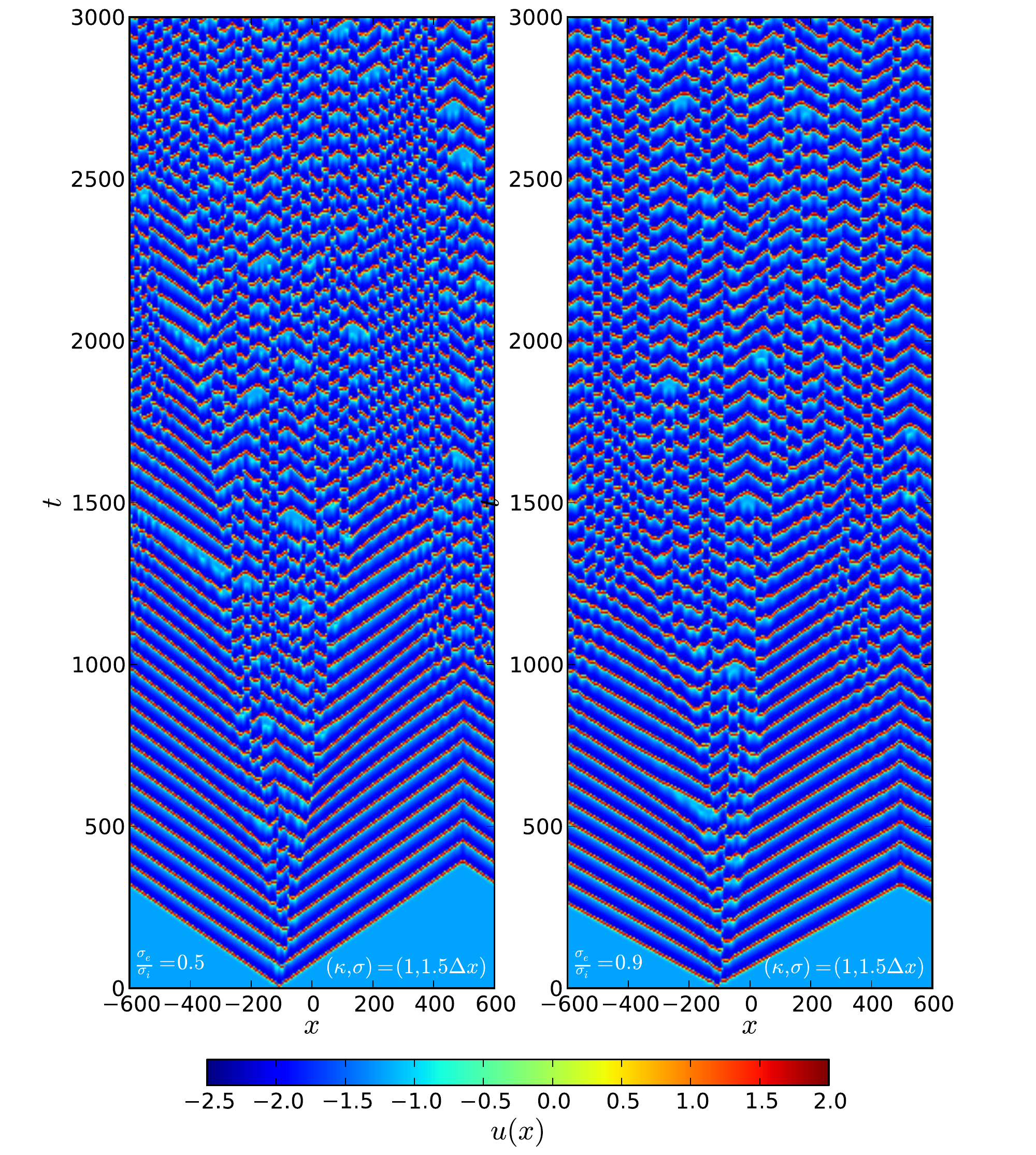}
		}
	\caption{Space-time plots with the Mexican hat kernel, $vv$-coupling-scheme, left: $r=\sigma_{e}/\sigma_{i}=0.5$, right: $r=\sigma_{e}/\sigma_{i}=0.9$. Other parameters: $\left(\kappa,\sigma_{i}\right)=\left(1,1.5\Delta x\right)$, $\left(\varepsilon,\beta\right)=\left(0.08,1.2\right)$, $L=1200$, integration timestep $dt=0.005$, spatial resolution $L/16384$. Initial condition: homogeneous steady state perturbed by uncorrelated random fluctuations.
 }
	\label{fig:mexhat_pattern}
\end{figure}

In Fig.~\ref{fig:mexhat} the ratio of the ranges of the short-range excitatory ($\sigma_{e}$) and the long-range inhibitory ($\sigma_{i}\equiv \sigma$) interaction $r=\sigma_{e}/\sigma_{i}=0.1$ is chosen as $r=0.1$. Changing the ratio $r$ in the interval $0<r<1$ has only quantitative effects on the $(\kappa,\sigma)$ control parameter plane, but retains the qualitative features. The larger $r$, the smaller are the $\sigma$ values at which the characteristic regions occur \cite{BAC13}. Thus, for instance, in the $uu$- and $vv$-self-coupling schemes the multiple pulse generation (MP) regimes extend to smaller $\sigma$, and in particular, in the $vv$-scheme the MP region, which is not visible in the ($\kappa, \sigma$)-range plotted in Fig.~\ref{fig:mexhat} for $r=0.1$, appears in the upper right corner for larger $r$.
This MP region needs to be highlighted, since it is not associated with a wave instability of the homogeneous steady states, as for the
$\delta$-function and exponential kernels. The resulting space-time patterns are more complex than simple wave patterns. Exemplary space-time plots are provided in Fig.~\ref{fig:mexhat_pattern} for $r=0.5$ and $r=0.9$. After two counterpropagating periodic wave trains are nucleated, they eventually lose their stability and end up in complex spatio-temporal patterns. 
The nucleation of periodic self-organized pacemakers is similar to what has been observed in three-variable excitable reaction-diffusion systems without nonlocal coupling \cite{STI09}.

\subsection{Complex spatio-temporal patterns}
\begin{figure}
        \centering
        \begin{subfigure}[b]{0.3\textwidth}
                \includegraphics[width=\textwidth]{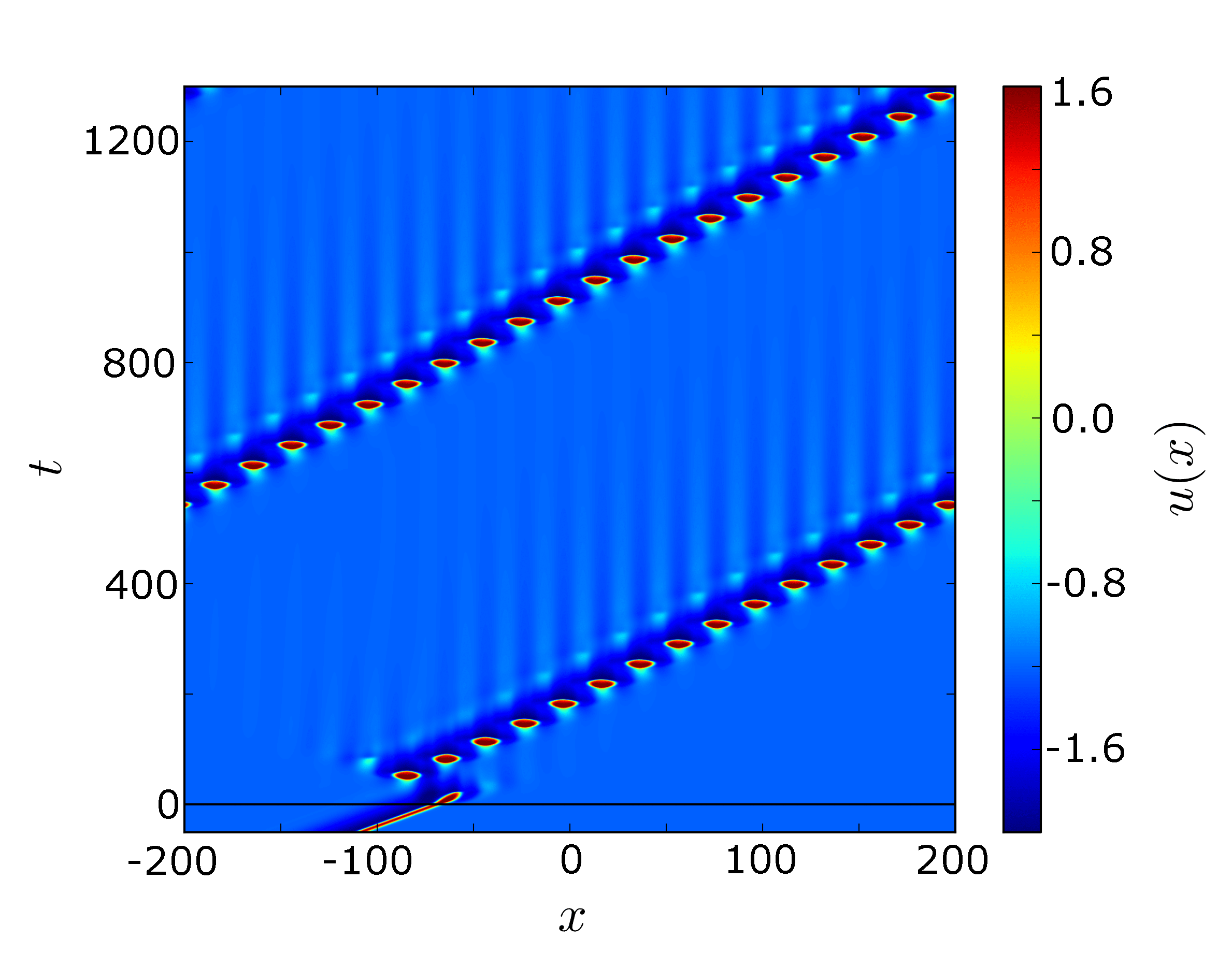}
								\caption{(a)}
                \label{fig:blinking_pulse}
        \end{subfigure}
         \begin{subfigure}[b]{0.3\textwidth}
                \includegraphics[width=\textwidth]{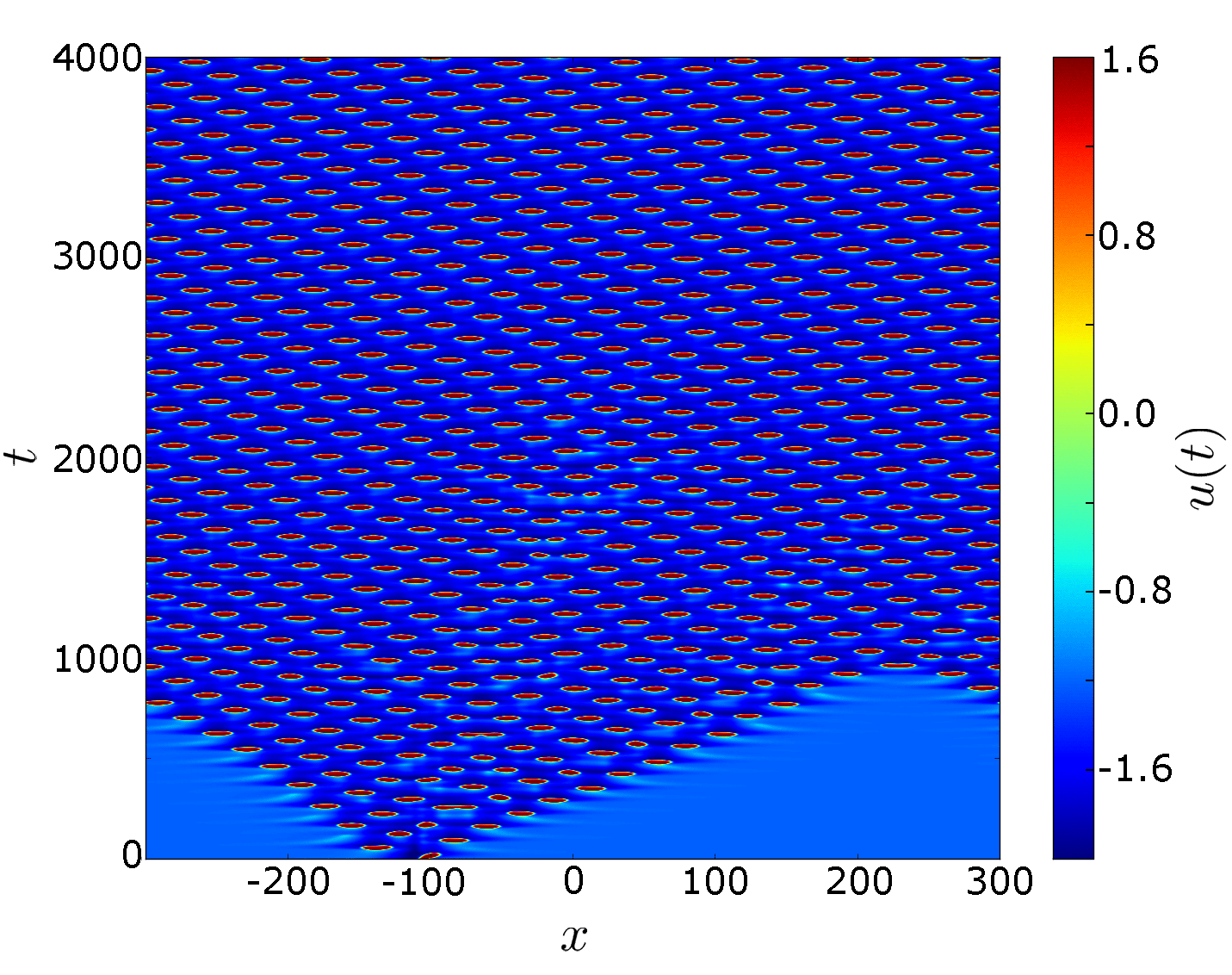}
								\caption{(b)}
                \label{fig:blinking_wave}
        \end{subfigure}
        \caption{Spatio-temporal plots of complex space time patterns. The initial condition is the stable pulse of the uncontrolled system. (a) Blinking traveling pulse: Isotropic $\delta$-function, $vu$-coupling scheme, $\left(\kappa,\sigma\right)=\left(0.48,1.16\Delta x\right)$. (b) Blinking traveling wave. Control configuration: exponential function, $vu$-coupling scheme, $\left(\kappa,\sigma\right)=\left(0.98,1.98\Delta x\right)$. 
Parameters: $\left(\varepsilon,\beta\right)=\left(0.08,1.2\right)$.
}
\label{fig:complex_spatio_temporal}
\end{figure}

In our simulations we have also observed other complex spatio-temporal patterns. In general, these occur for 
parameter values close to the boundary of the multiple pulse regime. Figure~\ref{fig:complex_spatio_temporal} 
shows examples of spatio-temporal plots of a blinking traveling pulse (a) and a blinking traveling wave (b), i.e.,
the propagating pattern is temporally modulated by switching the excitation periodically on and off. 
It appears that these patterns might be due to the interaction of Hopf and wave instabilities.

\section{Discussion}
\label{sec:discussion}
It has been shown that nonlocal control is able to accelerate, decelerate, and suppress an initially stable traveling pulse for various choices of the integral kernels and coupling schemes. Nonlocal control is also able to generate space-time patterns not present without control, e.g., Turing patterns and wave patterns, or even more complex patterns like salt-and-pepper instabilities or blinking pulses and waves. The control term can destabilize the homogeneous steady state in certain control parameter regimes. The boundaries of these instability regimes have been analytically described. 

In detail we have considered an isotropic $\delta$-function kernel, an exponential kernel, and a Mexican hat kernel, which combines short range activation with long range inhibition, and is thus of particular relevance for neuronal systems. As coupling schemes we have
used diagonal (activator-activator or inhibitor-inhibitor) and nondiagonal (activator-inhibitor or inhibitor-activator) coupling.
Activator self-coupling always gives rise to wave instabilities for the parameters investigated. The cross-coupling schemes and the  
inhibitor self-coupling scheme can cause Turing instabilities when using the Mexican hat control kernel, and salt-and-pepper instabilities when using the isotropic $\delta$-function or the exponential function. The spatial period of wave trains and Turing patterns can be tuned by changing the spatial coupling range of the control kernel.

The linear stability analysis of the homogeneous steady state also enables one to classify the type of the emerging instability as Turing, wave, or salt-and pepper instability. The analytical predictions for the stability boundaries of the homogeneous steady state are in good agreement with the simulation results of the nonlinear system. In the activator self-coupling scheme, the boundary of the regime of multiple pulse generation coincides precisely with the analytically calculated stability boundary of the homogeneous steady state. Care must, however, be taken for the other coupling schemes. There are control configurations that suppress or support stable pulses in the first place, but will eventually after long transient times destabilize the homogeneous steady state leading to the generation of pulses. Therefore in simulations sufficiently long simulation times must be used. 

In conclusion, a propagating pulse may be transformed into other spatio-temporal patterns by the nonlocal coupling, depending upon the parameters of the nonlocal control term. Since the obtained patterns depend sensitively upon these parameters, the choice of these parameters (coupling strength $\kappa$, coupling range $\sigma$, coupling scheme, and nonlocal coupling kernel) represents a convenient way of control of pulse propagation. We have systematically scanned the ($\sigma, \kappa$)-plane for each of the four coupling schemes ($uu, uv, vu, vv$) and different coupling kernels. Thus we have provided an exhaustive picture how nonlocal coupling affects pulse propagation in an excitable medium. For example, if pulse suppression is desired, the parameters should be chosen from the region denoted by PS in Figs.~\ref{fig:iso_delta}), \ref{fig:exp}, \ref{fig:mexhat}; and, analogously, if pulse acceleration or deceleration is desired (color coded regions), etc. It should also be noted from the figures that multistability between these different patterns can occur, e.g., between accelerated pulses (green) and salt-and-pepper instabilities (white dotted hatching), or between multiple pulse generation (dark grey, MP) and salt-and-pepper instabilities in Fig.~\ref{fig:iso_delta}, bottom right panel.

\section*{Acknowledgment}
This work was supported by DFG in the framework of SFB 910 ``Control of 
self-organizing nonlinear systems''. Helpful discussions with Julien Siebert are gratefully acknowledged.



\begin{thebibliography}{10}

\bibitem{HAK83}
H. Haken, {\em {Synergetics, An Introduction}}, 3 ed. (Springer, Berlin, 1983).

\bibitem{KUR84}
Y. Kuramoto, {\em Chemical Oscillations, Waves and Turbulence}
  (Springer-Verlag, Berlin, 1984).

\bibitem{MIK94}
A.~S. Mikhailov, {\em Foundations of Synergetics Vol. I}, 2 ed. (Springer,
  Berlin, 1994).

\bibitem{KAP95a}
{\em Chemical Waves and Patterns}, edited by R. Kapral and K. Showalter
  (Kluwer, Dordrecht, 1995).

\bibitem{KEE98}
J.~P. Keener and J. Sneyd, {\em Mathematical physiology} (Springer, New York,
  Berlin, 1998).

\bibitem{KAR13}
H. Karatas, S.~E. Erdener, Y. Gursoy-Ozdemir, S. Lule, E. Eren-Kocak, Z.~D.
  Sen, and T. Dalkara, Science {\bf 339},  1092  (2013).

\bibitem{DRE11}
J.~P. Dreier, Nat. Med. {\bf 17},  439  (2011).

\bibitem{DAH09a}
M.~A. Dahlem, R. Graf, A.~J. Strong, J.~P. Dreier, Y.~A. Dahlem, M. Sieber, W.
  Hanke, K. Podoll, and E. Sch{\"o}ll, Physica D {\bf 239},  889  (2010).

\bibitem{BEL81}
B.~N. Belintsev, M.~A. Livshits, and M.~V. Volkenstein, Z. Phys. B {\bf 44},
  345  (1981).

\bibitem{MAZ97}
N. Mazouz, G. Fl{\"a}tgen, and K. Krischer, Phys.~Rev.~E {\bf 55},  2260
  (1997).

\bibitem{SHE97a}
M. Sheintuch and O. Nekhamkina, J. Chem. Phys. {\bf 107},  8165  (1997).

\bibitem{KUR98}
Y. Kuramoto, D. Battogtokh, and H. Nakao, Phys. Rev. Lett. {\bf 81},  3543
  (1998).

\bibitem{HIL01}
M. Hildebrand, H. Sk\o{}dt, and K. Showalter, Phys. Rev. Lett. {\bf 87},
  088303  (2001).

\bibitem{LI01a}
Y.~J. Li, J. Oslonovitch, N. Mazouz, F. Plenge, K. Krischer, and G. Ertl,
  Science {\bf 291},  2395  (2001).

\bibitem{NIC02}
E.~M. Nicola, M. Or-Guil, W. Wolf, and M. B{\"a}r, Phys. Rev. E {\bf 65},
  055101  (2002).

\bibitem{VAR05}
H. Varela, C. Beta, A. Bonnefont, and K. Krischer, Phys. Rev. Lett. {\bf 94},
  174104  (2005).

\bibitem{NIC06}
E.~M. Nicola, M. B{\"a}r, and H. Engel, Phys.~Rev.~E {\bf 73},  066225  (2006).

\bibitem{BOR06}
G. Bordyougov and H. Engel, Phys.~Rev.~E {\bf 74},  016205  (2006).

\bibitem{GEL10}
L. Gelens, G. Gomila, G. Van~der Sande, and M.~A. Mat\'\i{}as, Phys. Rev. Lett.
  {\bf 104},  154101  (2010).

\bibitem{COL14}
P. Colet, M.~A. Mat\'\i{}as, L. Gelens, and D. Gomila, Phys. Rev.~E {\bf 89},
  012914  (2014), arXiv:1305.6801v1.

\bibitem{GEL14}
L. Gelens, M.~A. Mat\'\i{}as, D. Gomila, T. Dorissen, and P. Colet, Phys.
  Rev.~E {\bf 89},  012915  (2014).

\bibitem{LOE14}
J. L{\"o}ber, R. Coles, J. Siebert, H. Engel, and E. Sch{\"o}ll,  in {\em
  Engineering of Chemical Complexity {II}}, edited by A.~S. Mikhailov and G.
  Ertl (World Scientific, Singapore, 2014), arXiv:1403.3363.

\bibitem{PIS06}
L.~M. Pismen, {\em Patterns and Interfaces in Dissipative Dynamics}, {\em
  Springer Series in Synergetics} (Springer, Berlin, 2006).

\bibitem{PET94}
V. Petrov, M.~J. Crowley, and K. Showalter, Phys.~Rev.~Lett. {\bf 72},  2955
  (1994).

\bibitem{TAN03}
D. Tanaka and Y. Kuramoto, Phys. Rev. E {\bf 68},  026219  (2003).

\bibitem{SHI04}
S.-i. Shima and Y. Kuramoto, Phys. Rev.~E {\bf 69},  036213  (2004).

\bibitem{SIE14}
J. Siebert, S. Alonso, M. B{\"a}r, and E. Sch{\"o}ll, Phys. Rev.~E {\bf 89},
  052909  (2014).

\bibitem{MID92a}
U. Middya, M. Sheintuch, M.~D. Graham, and D. Luss, Physica D {\bf 63},  393
  (1992).

\bibitem{YOC10a}
A. Yochelis and M. Sheintuch, Phys. Rev. E {\bf 81},  025203  (2010).

\bibitem{ROV93}
A. Rovinsky and M. Menzinger, Phys. Rev. Lett. {\bf 70},  778  (1993).

\bibitem{KHA95b}
Y. Khazan and L.~M. Pismen, Phys. Rev. Lett. {\bf 75},  4318  (1995).

\bibitem{MAL96a}
H. Malchow, Journal of Marine Systems {\bf 7},  193  (1996), the Coastal Ocean
  in a Global Change Perspective.

\bibitem{HAR01a}
J. von Hardenberg, E. Meron, M. Shachak, and Y. Zarmi, Phys. Rev. Lett. {\bf
  87},  198101  (2001).

\bibitem{BAE94}
M. B{\"a}r, M. Falcke, M. Hildebrand, M. Neufeld, H. Engel, and M. Eiswirth,
  Int.~J.~Bifur.~Chaos {\bf 4},  499  (1994).

\bibitem{BET04a}
C. Beta, M.~G. Moula, A.~S. Mikhailov, H.~H. Rotermund, and G. Ertl, Phys. Rev.
  Lett. {\bf 93},  188302  (2004).

\bibitem{PLE01}
F. Plenge, P. Rodin, E. Sch{\"o}ll, and K. Krischer, Phys.~Rev.~E {\bf 64},
  056229  (2001).

\bibitem{MEI00b}
M. Meixner, P. Rodin, E. Sch{\"o}ll, and A. Wacker, Eur.~Phys.~J.~B {\bf 13},
  157  (2000).

\bibitem{SCH01}
E. Sch{\"o}ll, {\em Nonlinear spatio-temporal dynamics and chaos in
  semiconductors} (Cambridge University Press, Cambridge, 2001), {Nonlinear
  Science Series}, Vol. 10.

\bibitem{MIK06}
A.~S. Mikhailov and K. Showalter, Phys.~Rep. {\bf 425},  79  (2006).

\bibitem{CHR02a}
J. Christoph and M. Eiswirth, Chaos {\bf 12},  215  (2002).

\bibitem{KUR02a}
Y. Kuramoto and D. Battogtokh, Nonlin. Phen. in Complex Sys. {\bf 5},  380
  (2002).

\bibitem{ABR04}
D.~M. Abrams and S.~H. Strogatz, Phys.~Rev.~Lett. {\bf 93},  174102  (2004).

\bibitem{HAG12}
A.~M. Hagerstrom, T.~E. Murphy, R. Roy, P. H{\"o}vel, I. Omelchenko, and E.
  Sch{\"o}ll, Nature Physics {\bf 8},  658  (2012).

\bibitem{TIN12}
M.~R. Tinsley, S. Nkomo, and K. Showalter, Nature Physics {\bf 8},  662
  (2012).

\bibitem{MAR13}
E.~A. Martens, S. Thutupalli, A. Fourri{\`e}re, and O. Hallatschek, Proc. Nat.
  Acad. Sciences {\bf 110},  10563  (2013).

\bibitem{OME13}
I. Omelchenko, O.~E. Omel'chenko, P. H{\"o}vel, and E. Sch{\"o}ll, Phys. Rev.
  Lett. {\bf 110},  224101  (2013).

\bibitem{ZAK14}
A. Zakharova, M. Kapeller, and E. Sch{\"o}ll, Phys.~Rev.~Lett. {\bf 112},
  154101  (2014).

\bibitem{PAN14}
M.~J. Panaggio and D.~M. Abrams, arXiv:1403.6204  (2014).

\bibitem{KAN02}
K. Kang, M. Shelley, and H. Sompolinsky, PNAS {\bf 100},  2848  (2002).

\bibitem{XU04b}
X. Xu, W. Bosking, G. S\'{a}ry, J. Stefansic, D. Shima, and V. Casagrande, J.
  Neurosci. {\bf 24},  6237  (2004).

\bibitem{DAH08}
M.~A. Dahlem, F.~M. Schneider, and E. Sch{\"o}ll, Chaos {\bf 18},  026110
  (2008).

\bibitem{SCH09c}
F.~M. Schneider, E. Sch{\"o}ll, and M.~A. Dahlem, Chaos {\bf 19},  015110
  (2009).

\bibitem{DAH12b}
M.~A. Dahlem and T.~M. Isele, J. Math. Neurosci {\bf 3},  7  (2013).

\bibitem{KNE14}
F. Kneer, E. Sch{\"o}ll, and M.~A. Dahlem, New J.~Phys. {\bf 16},  053010
  (2014).

\bibitem{WIN87}
A.~T. Winfree, {\em When Time Breaks Down: The Three-Dimensional Dynamics of
  Electrochemical Waves and Cardiac Arrhythmias} (Princeton University Press,
  Princeton, NJ, 1987).

\bibitem{ALO03}
S. Alonso, F. Sagu{\'e}s, and A.~S. Mikhailov, Science {\bf 299},  1722
  (2003).

\bibitem{DAE13}
P. D{\"a}hmlow, S. Alonso, M. B{\"a}r, and M.~J.~B. Hauser, Phys. Rev. Lett.
  {\bf 110},  234102  (2013).

\bibitem{AHL07}
A. Ahlborn and U. Parlitz, Phys.~Rev.~E {\bf 75},  65202  (2007).

\bibitem{KYR09}
Y.~N. Kyrychko, K.~B. Blyuss, S.~J. Hogan, and E. Sch{\"o}ll, Chaos {\bf 19},
  043126  (2009).

\bibitem{GUR13b}
S.~V. Gurevich and R. Friedrich, Phys. Rev. Lett. {\bf 110},  014101  (2013).

\bibitem{STI13}
M. Stich and C. Beta, Phys. Rev. E {\bf 88},  042910  (2013).

\bibitem{HUT03}
A. Hutt, M. Bestehorn, and T. Wennekers, Network: Computation in Neural Systems
  {\bf 14},  351  (2003).

\bibitem{ZHA14}
L. Zhang and A. Hutt, Journal of Applied Analysis and Computation {\bf 4},  1
  (2014).

\bibitem{FIT61}
R. FitzHugh, Biophys. J. {\bf 1},  445  (1961).

\bibitem{NAG62}
J. Nagumo, S. Arimoto, and S. Yoshizawa., Proc. IRE {\bf 50},  2061  (1962).

\bibitem{LIN04}
B. Lindner, J. Garc{\'i}a-Ojalvo, A.~B. Neiman, and L. Schimansky-Geier,
  Phys.~Rep. {\bf 392},  321  (2004).

\bibitem{KON10}
S. Kondo and T. Miura, Science {\bf 329},  1616  (2010).

\bibitem{MEI97a}
M. Meixner, A. {De Wit}, S. Bose, and E. Sch{\"o}ll, Phys.~Rev.~E {\bf 55},
  6690  (1997).

\bibitem{BAC13}
C.~A. Bachmair, Nonlocal control of pulse propagation in excitable media, 2013,
  Master's Thesis, TU Berlin.

\bibitem{CRA06}
R.~V. Craster and R. Sassi, Technical Report {\bf 99},  1  (2006).

\bibitem{STI09}
M. Stich, A.~S. Mikhailov, and Y. Kuramoto, Phys. Rev. E {\bf 79},  026110
  (2009).

\end{thebibliography}

\end{document}